\documentclass[page-classic]{epl2} 
\usepackage{graphicx}
\usepackage{cite}
\usepackage{amssymb}

\title{Evidence of a multiple boson emission in Sm$_{1-x}$Th$_x$OFeAs}
\shorttitle{Multiple boson emission in Sm$_{1-x}$Th$_x$OFeAs} 

\author{S.A. Kuzmichev\inst{1,2} \and T.E. Kuzmicheva\inst{2} \and N.D. Zhigadlo\inst{3}}
\shortauthor{S.A. Kuzmichev \etal}

\institute{
\inst{1} Department of Low Temperature Physics and Superconductivity, M.V. Lomonosov Moscow State University, 119991 Moscow, Russia\\
\inst{2} Laboratory of Strongly Correlated Electron Systems, P.N. Lebedev Physical Institute, Russian Academy of Sciences, 119991 Moscow, Russia\\
\inst{3} Department of Chemistry and Biochemistry, University of Bern, Freiestrasse 3, CH-3012 Bern, Switzerland\\}

\abstract{We studied a reproducible fine structure observed in dynamic conductance spectra of Andreev arrays in Sm$_{1-x}$Th$_x$OFeAs superconductors with various thorium concentrations ($x = 0.08 \textendash 0.3$) and critical temperatures $T_c = 26\textendash50$\,K. This structure is unambiguously caused by a multiple boson emission (of the same energy) during the process of multiple Andreev reflections. The directly determined energy of the bosonic mode reaches $\varepsilon_0 = 14.8 \pm 2.2$\,meV for optimal compound. Within the studied range of $T_c$, this energy as well as the large $\Delta_L$ and the small $\Delta_S$ superconducting gaps, nearly scales with critical temperature with the characteristic ratio $\varepsilon_0/k_BT_c \approx 3.2$ (and $2\Delta_L/k_BT_c \approx 5.3$, correspondingly) resembling the expected energy $\Delta_L + \Delta_S$ of spin resonance and spectral density enhancement in $s^{\pm}$ and $s^{++}$ states, respectively.}

\begin{document}

\maketitle

Fe-based superconductors Sm$_{1-x}$Th$_x$OFeAs belong to the oxypnictide family (so called 1111), and have rather simple crystal structure, resembling the stack of superconducting FeAs blocks alternating with Sm$_{1-x}$Th$_x$O spacers along the $c$-direction \cite{Zhigadlo2010}. Under electron doping, the $T_c$ varies in the wide range, reaching 54\,K at $x \approx 0.3$ nominal concentration \cite{Zhigadlo2010,Zhigadlo2012}. Band-structure calculations \cite{Singh} showed the density of states at the Fermi level formed mainly by iron $3d$ states. For this reason, the (Sm,Th) substitution affecting the spacer structure barely seems not changing the underlying pairing mechanism \cite{PRB2017}. The Fermi surface consists of tubular sections, electron-like near the M point of the first Brillouin zone, and hole-like near the $\Gamma$ point, both with no significant $k_z$ anisotropy \cite{Singh,Charnukha}.

The majority of theoretical and experimental studies \cite{Singh,PRB2017,Charnukha,Si,Johnston,Borisenko} suppose two superconducting condensates developing below $T_c$. Earlier we reported the scaling between both gaps ($\Delta_L$ --- large gap, $\Delta_S$ --- small gap) and $T_c$, keeping $2\Delta_S/k_BT_c \approx 1.2-1.6$ and $2\Delta_L/k_BT_c = 5.0-5.7$ \cite{JETPL2014,UFN2014,JSNM2016,PRB2017}. Similar $2\Delta_L/k_BT_c$ was obtained in literature for Sm-1111 in point-contact probes \cite{Daghero_Sm,Wang}, and for various other 1111 \cite{Tanaka,Samuely,LOFA,Mukuda,Noat,Prakash,Matano}. Both BCS-ratios diverge from the weak-coupling BCS prediction due to a strong coupling in the ``driving'' bands where the large gap is developed, and a $k$-space proximity effect with the ``driven'' $\Delta_S$ bands. The pairing mechanism in Fe-based superconductors is still puzzling. Three basic models, $s^{++}$, $s^{\pm}$, and shape resonance model, were proposed so far \cite{Si,Mazin,Korshunov2,Hirschfeld,Hirschfeld2011,Onari,Kontani,Bianconi}. A sharp peak in the imaginary spin susceptibility appearing at nesting vector and a certain energy, is the signature of $s^{\pm}$ mechanism mediated by spin fluctuations \cite{Eremin,Maier}. A number of neutron diffraction studies reported a clear ``magnetic resonance'' peak, which energy roughly scales with $T_c$ \cite{Dai,Paglione}. In $s^{++}$ approach, imaginary part of dynamic spin susceptibility demonstrates a smeared maximum rather than peak \cite{Korshunov2,Onari,Kontani}. More recent theoretical studies showed that in framework of both $s^{++}$ and $s^{\pm}$ models the dynamic spin susceptibility has a feature near $\Delta_L + \Delta_S$ energy, a sharp peak related to spin resonance in $s^{\pm}$ state, or spectral density enhancement above $\Delta_L + \Delta_S$ in $s^{++}$ state \cite{Korshunov,Onari}.

Tunneling contact probes could provide information about electron-boson interaction \cite{BJ,Zimmermann,Pon_THz,Daghero,Tortello,Tortello_Ba,Pon_Legg1,Pon_Legg2,PCS_book,Fasano,Shan}. For tunneling normal metal --- insulator --- superconductor (NIS) junction, the derivative d$^2$I/dV$^2$ of dynamic conductance spectrum at bias voltages $V > \Delta/e$ represents a spectral function of electron-boson interaction \cite{PCS_book}. In other words, the edge energy $\Delta$ changes into $(\Delta + \varepsilon_0)$ for some electrons, where $\varepsilon_0 < 2\Delta$ is a particular boson energy. In Nd-1111, the Eliashberg function extracted from d$^2$I(V)/dV$^2$ \cite{Tanaka} well matches the calculated one, and the phonon density of states \cite{LeTacon}. Scanning tunneling spectroscopy (STS) studies with Ba$_{0.6}$K$_{0.4}$Fe$_2$As$_2$ revealed a fine structure attributed to a coupling of quasiparticles with a bosonic mode near 14\,meV \cite{Shan}. In contrast, in STS probe with SmFeAsO$_{1-x}$F$_x$ \cite{Fasano}, the minimum of the dip-hump structure was attributed to a sign of a resonance spin mode within the energy range $\varepsilon_{res} = 2-8$\,meV. A strange correlation was detected in \cite{Fasano}: $\varepsilon_{res} + \Delta = 11-12~{\rm meV} = const$. Some studies of NS point contacts in nearly optimal F-substituted SmFeAsO$_{0.8}$F$_{0.2}$ reported a dynamic conductance fine structure observed above the gap edge and therefore attributed with electron-boson interaction. The complex shape of the dI(V)/dV spectra \cite{Daghero,Tortello} were fitted using three-gap model solely, the largest gap had the BCS ratio 8.7. A close BCS-ratio and resembling fine structure were observed in BaFe$_{1.8}$Co$_{0.2}$As$_2$ \cite{Daghero,Tortello,Tortello_Ba}. A clear maximum offset the largest gap bias was interpreted as a manifestation of interaction of electrons with a bosonic mode with the energy $\varepsilon_0 \approx 22$\,meV for Sm-1111. However, the energy rapidly decreased with temperature in a gap-like way and therefore seemed to have non-phononic origin. Similar looking high-bias features often emerge in point contact probes of 1111 oxypnictides \cite{Samuely,Chen,Naidyuk,Yates}, nonetheless, those fine structures were not assigned any physical meaning there. By contrast, in \cite{Chen} it was pointed out that the position of those features varied with respect to contact resistance, thus doubting their essentiality. Generally speaking, there are difficulties in the interpretation of dI(V)/dV spectra of point contacts using three-gap approach due to abundance of fitting parameters (up to 11) \cite{Daghero}, and necessarily accounting the partial spectral Eliashberg functions (for each band). Thus, the latter remains unsolved experimental issue. If the gap function $\Delta_i(\omega)$ has nonmonotonic features, the dynamic conductance spectrum would also show features above the main gap bias voltage ($2\Delta/e$ for SnS-contact, and $\Delta/e$ for N(I)S contact). Fortunately, probing SnS-contact, it is possible to distinguish between the boson-caused resonance and the above mentioned $\Delta(\omega)$-caused features, because the latter do not generate a subharmonic structure (see below).

In SnS contact, normal electron could emit a boson with energy less than $2\Delta$ during the process of multiple Andreev reflections. In our recent studies of nearly optimal GdO$_{1-x}$F$_x$FeAs oxypnictides with critical temperatures $T_c = 46 - 50$\,K, we observed the reproducible fine structure caused by electron-boson interaction \cite{JETPL2017}. We unambiguously showed the fine structure was caused by the bulk, and position of the bosonic resonances do not depend on the contact area and resistance. The directly determined energy of the bosonic mode $\varepsilon_0 = 11 \pm 2$\,meV did not exceed $2\Delta_L$ edge value, and correlated with $\Delta_L + \Delta_S$. Here we present a study of a fine structure in dynamic conductance spectra of Andreev arrays in thorium-substituted Sm$_{1-x}$Th$_x$OFeAs with $x = 0.08 - 0.3$ and critical temperatures $T_c = 26-50$\,K. For optimal compound, the energy of the characteristic bosonic mode reaches $\varepsilon_0 = 14.8 \pm 2.2$\,meV (or $118 \pm 18$\,cm$^{-1}$), and scales with critical temperature, keeping nearly constant ratio $\varepsilon_0/k_BT_c \approx 3.2$. The latter is close to the expected position of the resonance peak of imaginary spin susceptibility in $s^{\pm}$ state \cite{Korshunov} or enhanced spectral density peak in $s^{++}$ state \cite{Kontani}.

Polycrystalline Sm$_{1-x}$Th$_x$OFeAs samples with various thorium doping were synthesized by high-pressure method. Overall details of the sample cell assembly and high-pressure synthesis process may be found in \cite{Zhigadlo2010,Zhigadlo2012}. Powders of SmAs, ThAs, Fe$_2$O$_3$, and Fe of high purity ($\geq 99.95 \%$)
were weighed according to the stoichiometric ratio, thoroughly ground, and pressed into pellets. Then, the pellet containing precursor was enclosed in a boron nitride crucible and placed inside a pyrophyllite cube with a graphite heater. All the preparatory steps were done in a glove box under argon atmosphere. The six tungsten carbide anvils generated pressure on the whole assembly. In a typical run, the sample was compressed to 3\,GPa at room temperature. While keeping the pressure constant, the temperature was ramped up within 1\,h to the maximum value of 1430\,$^{\circ}$C, maintained for 4.5\,h, and finally quenched to the room temperature. Afterward, the pressure was released and the sample removed. Subsequently recorded X-ray powder diffraction patterns revealed high homogeneity of the samples and the presence of a single superconducting phase \cite{Zhigadlo2010}. The amount of additional nonsuperconducting phases SmAs and ThO$_2$ was vanishingly small. The bulk character of superconductivity in Sm$_{1-x}$Th$_x$OFeAs samples was confirmed by magnetization measurements.

In order to form SnS-Andreev contact, we used a ``break-junction'' technique. More details about our set up could be found elsewhere \cite{BJ,Moreland}. A plate-like sample was attached onto a springy sample holder and cooled down to $T = 4.2$\,K. Then, the holder was gently curved, thus cracking the crystal. Two cryogenic surfaces coupled with a weak link were kept in the bulk of the sample during the studies. We did not separate the clefts to a moderate distance, facilitating clean and non degraded cryogenic surfaces \cite{BJ}. In Sm-1111, the weak link formally acts as thin normal metal \cite{JETPL2014,UFN2014,JSNM2016,PRB2017}, as the resulting I(V) and dI(V)/dV resemble those of a clean classical SnS-contact \cite{OTBK,Arnold,Averin,Kummel}.

Steps and terraces commonly appear on cryogenic clefts and may realize SnSn-\dots-S arrays typical for the break-junction studies of single crystals and even polycrystalline samples of layered compounds \cite{BJ,PRB2017}. Layered grain splits when making the crack, and its $ab$ crystallographic plane oriented nearly parallel to the crack, and shows steps and terraces likewise in single crystal \cite{BJ,PRB2017}. The array is a stack of $m$ identical SnS junctions along the $c$-direction. Tuning the curvature of the holder makes the terraces slide along the $ab$-planes, forming SnS junctions and arrays with various area and $m$. With this set up, one could probe dozens of Andreev contacts in one and the same sample, and collect reproducible and self-consistent data.

Multiple Andreev reflection effect (MARE) occurring in ballistic SnS contact with a constriction narrower than carrier mean free path \cite{Sharvin}, causes a pronounced excess current (``foot'') near zero-bias region in current-voltage characteristic (CVC), and a subharmonic gap structure (SGS) --- a sequence of dynamic conductance dips (in case of transparency of NS interfaces as high as 95 - 98\,\%).
Their positions $V_n = 2\Delta/en$, $n$ is natural subharmonic order \cite{OTBK,Arnold,Averin,Kummel}, directly determine the value of superconducting order parameter at any temperatures up to $T_c$ \cite{OTBK,Kummel}. The first Andreev minimum could be shifted towards zero for several reasons \cite{BJ,Averin,Kummel}; if it is the case (see Figs. 1-4), the gap value is determined using the positions of high-order subharmonics. In two-gap superconductor, two sets of dI(V)/dV features corresponding to the large and the small gap should be observed. The contact area typical for Sm-1111, is about $10 - 30$\,nm, as estimated \cite{BJ}, thus providing local measurements of energy parameters. Intrinsic MARE (IMARE) similar to intrinsic Josephson effect \cite{Pon_IJE} takes place in Andreev arrays and scales by a factor of $m$ the position of any features caused by the bulk. In particular, SGS's would appear at bias voltages $V_n = 2m \times \Delta_{L,S}/en$. The actual number of junctions in the array could be determined when normalize the dI(V)/dV by a factor of the natural $m$, until the positions of the main conductance features would coincide with those in the spectrum of single SnS-junction \cite{BJ,PRB2017}. The IMARE spectroscopy of SnSn-\dots-S break-junctions is therefore a direct local probe providing a highly accurate bulk values of characteristic energy parameters \cite{BJ}.

When undergoes (I)MARE, an electron could emit a boson with the energy $\varepsilon_0$ up to $2\Delta$. Boson absorption is nearly impossible, due to the lack of excited bosons at low temperatures. When the bosonic mode has a particular energy $\varepsilon_0$, one should observe satellite dips beyond the SGS at bias voltages \cite{Zimmermann,Pon_Legg1,Pon_Legg2,JETPL2017}
\begin{equation}
V_{n,k} = \frac{2\Delta + k\varepsilon_0}{en},
\end{equation}
($k$ is a natural number of sequentially emitted bosons). Since the amount of electrons emitting a boson decreases with $k$ increasing, the satellites are less pronounced. In case of sequential bosons emitted $k > 1$, $k$ equidistant satellites with diminishing intensity (due to $\Gamma$ broadening) would follow each gap subharmonic. One should not expect any $(k + 1)$-order feature if $k$'th dip became smeared. The bosonic energy could be directly determined as a ``distance'' between $2\Delta$ and $(2\Delta + \varepsilon_0)$ dips. When $\varepsilon_0$ small compare to $\Delta$ the bosonic features are identified unambiguously, since appear next to the Andreev minimum and almost not superpose with the SGS dips \cite{JETPL2017}. However, in case of $\varepsilon_0 \sim \Delta$, the satellites are located far from the ``parent'' SGS dips, and may overlap with the high-order subharmonics ($n \geq 2$). In general, merging dips may intensify the resulting conductance feature, likewise interfering.

The ``break-junction'' technique is a universal probe of superconducting order parameter and electron-boson interaction \cite{BJ}. It provides high quality of the contacts even in polycrystalline samples of layered compounds \cite{BJ}, giving opportunity to resolve a clear fine structure accompanying SGS in dynamic conductance spectra. Satellite structure at bias voltages corresponding to Eq. (1) was firstly observed in dI(V)/dV of microwave irradiated SnS-Andreev break junctions in YBaCuO \cite{Zimmermann}. Later, in Mg(Al)B$_2$, Ponomarev et al. \cite{Pon_Legg1,Pon_Legg2} reproducibly observed up to 4 satellites accompanied the large gap subharmonics ($2\Delta_{\sigma}/en$). The bosonic mode was interpreted there as Leggett plasma mode with maximum energy $\omega_L = 4-5$\,meV for undoped magnesium diborides. Similar energy was obtained in tunneling SIS contact studies, extracted from a fine structure caused by a resonant excitation of Leggett plasmons mode by Josephson supercurrent \cite{Pon_Legg1,Pon_Legg2}. According to theory \cite{Leggett}, that energy was not exceeded the doubled small gap, and evolved as $\omega_L^2 \sim \Delta_{\sigma} \cdot \Delta_{\pi}$ within nearly full range of aluminum concentration and $T_c = 6 - 41$\,K \cite{Pon_Legg1,Pon_Legg2}. Here in Sm-1111, we reproducibly observe up to $k = 4$ equidistant satellites, which evidences their bosonic origin.

\label:{Fig.1} shows normalized CVC (blue line, left vertical scale) measured at $T = 4.2$\,K of Andreev array in Sm-1111 sample with nearly optimal thorium concentration $x = 0.3$ and critical temperature $T_c \approx 49$\,K. The CVC is symmetric and non-hysteretic, and has a pronounced foot area with a significant excess current at low biases, typical for high-transparent SnS-contact. The contact resistance $R \approx 20 {\rm \Omega}$ (per one SnS-junction) is comparatively large indicating a ballistic transport. Taking the average product of bulk resistivity and carrier mean free path $\rho l^{el} \approx 5 \times 10^{-10}$\,$\Omega \cdot \rm{cm^2}$ nearly constant for Sm-1111 \cite{Tropeano1,Tropeano2}, and $\rho \approx 0.09$\,m$\Omega \cdot \rm{cm}$ for the optimal single crystal from the same batch \cite{Zhigadlo2010}, we use Sharvin formula $R = \frac{4}{3\pi}\frac{\rho l}{a^2}$ \cite{Sharvin}, and get the contact dimension $a \approx 33$\,nm which is less than $l^{el} \approx 55$\,nm. We note that for the experimental observation of MARE namely $l^{in}/2a$ ratio is essential ($l^{in}$ --- inelastic mean free path). Usually, $l^{in}$ is several times larger than $l^{el}$ facilitating the ballistic regime.

\begin{figure}
\includegraphics[width=.49\textwidth]{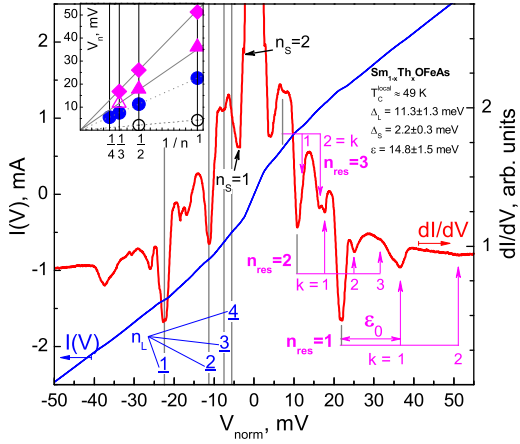}
\caption{Current-voltage characteristic (blue line, left vertical scale), and dI(V)/dV spectrum (red line, right scale) for Andreev array in nearly optimal Sm$_{0.7}$Th$_{0.3}$OFeAs sample with $T_c \approx 49$\,K. SGS of the large gap $\Delta_L \approx 11.3$\,meV is shown by gray vertical ticks and $n_L$ labels, for the small gap $\Delta_S \approx 2.2$\,meV \textemdash by black arrows and $n_S$ labels, the bosonic features with the energy $\varepsilon_0 \approx 14.8$\,meV are labelled with $n_{res}$ and vertical magenta arrows. The inset shows the positions of the $\Delta_L$ (blue circles), $\Delta_S$ (open circles), and bosonic features (triangles for $k=1$, rhombs for $k=2$ bosons emitted) versus the inverse number $1/n$. Gray lines are guidelines.}
\label{fig.1}
\end{figure}

\begin{figure}
\includegraphics[width=.49\textwidth]{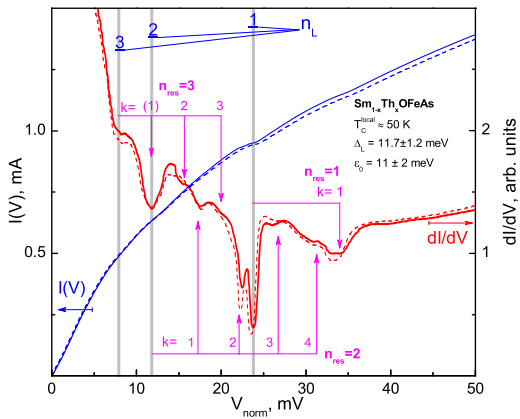}
\caption{Current-voltage characteristic (blue line, left vertical scale), and dynamic conductance spectrum (red line, right scale) for Andreev array in nearly optimal Sm$_{0.7}$Th$_{0.3}$OFeAs sample with $T_c \approx 50$\,K. Subharmonic gap structure of the large gap $\Delta_L \approx 11.7$\,meV is shown by gray vertical bars and $n_L$ labels. The features caused by the boson emission with the energy $\varepsilon_0 \approx 11$\,meV are labelled with $n_{res}$ and arrows. Dashed lines show the inverse negative parts of I(V) and dI(V)/dV.}
\label{fig.2}
\end{figure}

The dynamic conductance spectrum (red line, right scale in \label:{Fig.1}) demonstrates four Andreev features of the large gap (marked with $n_L = 1 - 4$ and gray vertical lines) located at $|V_n| \approx 23, 11.3, 7.5, 5.6$\,mV. The positions $V_n$ depend linearly on their inverse number $1/n$ (blue circles in the inset), thus composing the large gap SGS and directly determine the magnitude $\Delta_L \approx 11.3$\,meV. The dips more intensive than $n_L = 4$ and located at $\pm 4.1$\,mV do not satisfy the expected positions of 5$^{\rm th}$ subharmonic of the large gap. These features, and those observed at $\pm 2.2$\,mV, are obviously compose the second SGS related to the small gap $\Delta_S \approx 2.2$\,meV (open circles in the inset). The obtained gap values are in good agreement with the earlier IMARE studies \cite{UFN2014,JSNM2016,PRB2017}, and resemble those in sister compounds GdO$_{1-x}$F$_x$FeAs with similar $T_c$ \cite{EPL,UFN2014,JETPL2017}.

A rich fine structure resolved in the dI(V)/dV is labeled with $n_{res} = 1, 2, 3$ in \label:{Fig.1}. Next to the main harmonic $n_L = 1$ of the large gap, the clearly visible satellite is located at $eV_{res1} = 2\Delta_L + \varepsilon \approx 36$\,meV corresponding to a single $k = 1$ boson emitted by normal carriers. While, at the expected position of $k = 2$ resonance (two sequentially emitted bosons), $eV_{res1}(k=2) = 2\Delta_L + 2\varepsilon_0 \approx 51$\,meV, we observed only smeared feature. In the majority of obtained dI(V)/dV, the $(n = 1, k > 1)$ peculiarities following $2\Delta_L$-dips are hardly observable. The reason for this lies in the short propagation time $t_{cross}$ for the carriers driven by relatively high bias. In ballistic regime applicable to our constrictions, $t_{cross} \sim 1/V^2$, suggesting nearly no time for resonant energy transmission for $eV > 4\Delta_L$. At the half of these biases, $V_{res2} = V_{res1}/2 \approx \pm 18.7$\,mV the second boson-caused subharmonic ($n = 2$, $k =1$) should appear in accordance with Eq. (1). This position matches the external minimum of the doublet observed between the large gap subharmonics $n_L = 1, 2$. Indeed, the spectrum shows $k = 2$ dips at $\pm 26.2$\,mV and $k = 3$ features of a vanishing amplitude at $\pm 33.6$\,mV. These minima are nearly equidistant (see the magenta arrows in \label:{Fig.1}) and offset by $\approx 7.4$\,mV. This shift is exactly twice smaller than the distance $|eV_{res1} - 2\Delta_L| \equiv \varepsilon_0$. The $n = 3$, $k = 1$ bosonic feature is unresolved as a distinct dip since its expected position (open triangle in the inset) nearly matches the $\Delta_L/e$. Nonetheless, the $n = 3$, $k = 2$ feature corresponds to the internal minimum in the doublet at $\pm 17$\,mV. The bosonic resonances accompanying $n_L = 4$ are hardly resolvable due to minor intensity of the gap features. Overall, the fine structure features observed in the dI(V)/dV could be interpreted as boson-caused since satisfy Eq. (1). For certain $k$, their positions comprise a distinct subharmonic structure (triangles in the inset for $k = 1$, rhombs for $k = 2$). With $k$ increasing, the satellites smearing. Taking into account that some bosonic resonances merging with the large gap dips, we observe gradual (with no missed $k$ numbers) ``comb'' of satellites accompanying the $n = 2, 3$ SGS dips. According to formula (1), the energy of the bosonic mode could be directly determined as $\varepsilon_0 = \langle en|V^{res}_{n,k} - V^L_n|/k \rangle = 14.8$\,meV.

\begin{figure}
\includegraphics[width=.49\textwidth]{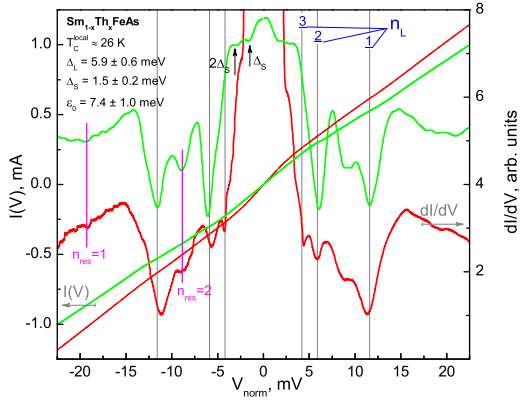}
\caption{I(V) (left scale), and dynamic conductance (right scale, corresponding colors) for two Andreev arrays in underdoped Sm$_{1-x}$Th$_x$OFeAs with $x < 0.08$ and $T_c \approx 26$\,K. SGS of the large gap $\Delta_L \approx 5.9$\,meV is shown by gray vertical lines and $n_L$ labels, for the small gap $\Delta_S \approx 1.5$\,meV --- by the black arrows, the features caused by boson emission with the energy $\varepsilon_0 \approx 7.4$\,meV are labelled with $n_{res}$ and magenta vertical bars.}
\label{fig.3}
\end{figure}

Similar fine structure corresponding up to $k = 4$ emitted bosons was observed in another sample from the same batch. \label:{Fig.2} shows normalized CVC (blue line, left scale), and dynamic conductance (red line, right scale) of Andreev array with $T_c = 50$\,K. The reversed negative voltage parts of I(V) and dI(V)/dV (dashed lines) show high symmetry of the characteristics. In the spectrum, the small gap SGS is invisible (which is typical for IMARE studies of optimally doped Sm-1111 \cite{PRB2017}), whereas three subharmonics of the large gap $\Delta_L \approx 11.7$\,meV are clearly seen (gray vertical lines and $n_L$ labels in \label:{Fig.2}). As in \label:{Fig.1}, we observe the single satellite offset $2\Delta_L/e$ bias. Herewith, the higher-order gap subharmonics are accompanied with multiple (sequential) bosonic resonances (arrows, $n_{res} = 2, 3$, and $k = 1 - 4$ labels). The doublet shape of the main SGS dip ($n_L = 1$) is not reproducible for other subharmonics therefore results from a $n = 2$, $k = 2$ dip. All the dynamic conductance features satisfy to Eq. (1). The resulting energy of the bosonic mode is the same order of magnitude, although a bit lower $\varepsilon_0 = 11 \pm 2$\,meV.

In underdoped Sm-1111 with critical temperature $T_c \approx 26$\,K the obtained Andreev spectra demonstrate the main boson dips with $n,k = 1$ barely (\label:{Fig.3}), probably relating with enhanced smearing $\Gamma$. The first feature $n_{res} = 1$ is located at $|V| \approx 19.3$\,mV and approximately offset by $\varepsilon_0/e \approx 7.4$\,mV the $2\Delta_L$ dip. The second feature at $8.8$\,mV is observed between $n_L = 1, 2$ subharmonics of the large gap $\Delta_L \approx 5.9$\,meV. The third feature expected at $V_{res3}\approx 6.7$\,mV seems unobservable due to smeared ``parent'' dips $n_L = 3$. Note despite the $n_L = 1$ fragment comprising the two dips at $\approx 11.6$ and $\approx 8.8$\,mV visually resembles the doublet typical for a case of four-fold gap distribution in $k$-space (see \cite{BJ}), the other $\Delta_L$ subharmonics are not doublet-like. Therefore, the large gap anisotropy cannot be a reason for the observed fine structure.

In order to compare the positions of the bosonic satellites relatively to the large gap SGS, in \label:{Fig.4} we show the fragments of dI(V)/dV spectra at $T = 4.2$\,K normalized by a value of $\Delta_L$. The characteristics were obtained in Andreev arrays of Sm-1111 samples with various thorium concentrations $x = 0.08 - 0.3$ and corresponding critical temperatures $T_c = 26 - 50$\,K. The upper spectrum is taken from \label:{Fig.1}. The lower spectrum reproduces the upper curve in \label:{Fig.3} (the fragment comprising the $n_{res} = 1$ bosonic feature was stretched vertically for clarity). Clearly, within the significant $T_c$ variation, the positions of the bosonic resonances for $n_{res} = 1, 2$, $k = 1$ (arrows in \label:{Fig.4}) are in a good agreement.

\begin{figure}
\includegraphics[width=.49\textwidth]{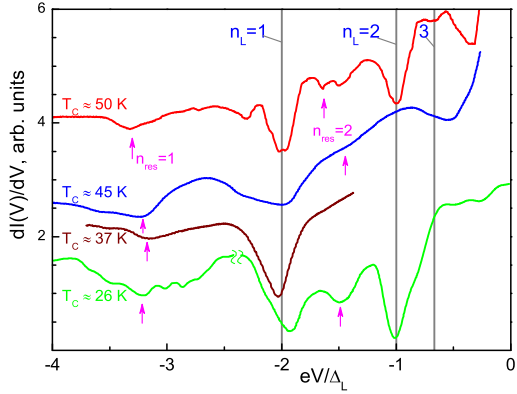}
\caption{Dynamic conductance spectra for Andreev arrays in Sm$_{1-x}$Th$_x$OFeAs samples with various thorium concentration and $T_c = 26 - 50$\,K. The dI(V)/dV are normalized with $\Delta_L$. SGS of the large gap is shown by gray vertical lines and $n_L$ labels. The $k=1$ bosonic features are labelled with $n_{res}$ and arrows. The fragment of the lower dI(V)/dV comprising the $n_{res}=1$ feature was vertically stretched for clarity.}
\label{fig.4}
\end{figure}

The summary of the data presented as follows:

(1) A ``comb'' of up to 4 equidistant satellites accompanying the $\Delta_L$ subharmonics is observed in dI(V)/dV spectra of Andreev arrays. The satellites are located in agreement with Eq. (1) and therefore seem to have electron-boson origin. Excepting those merging with SGS dips, the intensity of the satellites decreases with $k$ increase.

(2) This effect and the corresponding comb structure obviously has a bulk origin since observed during IMARE in SnS-arrays; the bias voltages $V_{res}$ scale with the number of junctions $m$ in array, together with both gaps SGS's. However, the satellites are less pronounced as compared with $\Delta_L$ dips, therefore, just a portion of carriers undergoing Andreev reflections emit a boson(s).

(3) The position of the satellites well correspond for various Andreev arrays, does not depend on the contact area and resistance, thus cannot be attributed as an artifact or caused by any dimensional effect.

(4) The observed fine structure do not match neither $2\Delta_{L,S}/en$ nor $(\Delta_L + \Delta_S)/en$ subharmonic sequence. Neither any of the satellites can relate to a distinct, the largest order parameter. In the case, it would have the BCS ratio $2\Delta_3/k_BT_c > 8$. Although agrees with PCAR results with fluorine-doped Sm-1111 \cite{Daghero}, the presence of three distinct gaps was not confirmed unambiguously neither theoretically nor experimentally for oxypnictide family (for a review, see \cite{Johnston,UFN2014,PRB2017,Si}). In addition, our preliminary data show the temperature behaviour of this fine structure does not resemble the expected $\Delta(T)$. For this reason, we cannot attribute the satellites with a $\Delta_L$ anisotropy in the $k$-space. However, this issue requires further studies.

(5) In the \label:{table}, we present the directly determined (using the data in \label:{Fig.1} -- \label:{Fig.4}, no fitting is needed \cite{OTBK,Kummel}) energy parameters of Sm-1111 within $T_c = 26 - 50$\,K. For optimal compound, $\varepsilon_0$ is up to $15$\,meV and agrees well with that determined for a sister compounds GdO$_{1-x}$F$_x$FeAs with similar $T_c$ \cite{JETPL2017}. For the entire $T_c$ range, the experimental value of $\varepsilon_0$ obviously do not exceed $2\Delta_L$, thus do not violate the MARE regime condition. Although carriers from each band undergo MAR, the boson emission is prohibited for normal carriers from the $\Delta_S$-band(s) due to $\varepsilon_0 > 2\Delta_S$. This is the reason why the bosonic satellites are observed next to the $\Delta_L$ subharmonics barely. Together with the large and the small gaps \cite{UFN2014,PRB2017}, the bosonic energy roughly scales with critical temperature, evidencing the emitted bosons are not phonons. Despite for the lowest $T_c \sim 26$\,K, $\varepsilon_0$ meets the lowest-frequency optic phonon mode $\hbar\omega_{phon} = 11- 14$\,meV (determined in Raman spectroscopy \cite{Zhao}, inelastic neutron and X-ray scattering studies \cite{LeTacon,Christianson} of various 1111), the latter remains nearly constant rather than scales with $T_c$ decrease.

(6) Unlike magnesium diborides \cite{Pon_Legg1,Pon_Legg2}, one cannot attribute the observed bosonic mode as Leggett plasma mode \cite{Leggett}. Firstly, several theoretical studies shown that Leggett plasmons are unobservable in iron pnictides \cite{Burnell,Ota}. Secondly, $\varepsilon_0^2 \nsim \Delta_L \cdot \Delta_S$ within the studied range.

(7) Instead, $\varepsilon_0 \approx \Delta_L + \Delta_S$ (see \label:{table}), and resembles the energy of spin resonance peak in $s^{\pm}$ state as predicted in \cite{Korshunov} or the enhanced spectral peak in $s^{++}$ state \cite{Onari}. For the bosonic mode, the average characteristic ratio is $\varepsilon_0/k_BT_c \approx 3.2$ (see \label:{table}). However, it should not be confused with weak-coupling limit of the BCS theory, since $\varepsilon_0$ does not represent the Cooper pair self-energy for any condensate. Note, for the ``leading'' large superconducting gap $\Delta_L$, the characteristic ratio well exceeds the BCS limit for the studied samples: $2\Delta_L/k_B T_c \approx 5.3$.

In conclusion, we have studied a fine structure reproducibly observed in dynamic conductance spectra of Andreev arrays in Sm$_{1-x}$Th$_x$OFeAs oxypnictides with thorium concentrations $x = 0.08 - 0.3$ and corresponding critical temperatures $T_c = 26 - 50$\,K. We unambiguously show that this structure is caused by a resonant sequential boson emission during IMARE. The directly determined energy of the bosonic mode scales with critical temperature together with $\Delta_L$ and $\Delta_S$. At $T_c \sim 50$\,K, $\varepsilon_0$ reaches $\approx 15$\,meV ($120 \pm 20$\,cm$^{-1}$) and resembles that determined by us earlier for GdO$_{1-x}$F$_x$FeAs with similar $T_c$). One cannot attribute the observed bosonic resonance with Leggett mode or optic phonon mode, nonetheless the $\varepsilon_0$ is close to the expected position of the the energy of spin resonance peak in $s^{\pm}$ state \cite{Korshunov} or the enhanced spectral density maximum in $s^{++}$ state \cite{Onari}.

\begin{table}
\caption{The superconducting gaps and the energy of the bosonic mode directly determined for Sm$_{1-x}$Th$_x$OFeAs.}
\begin{tabular}{cccccc}
\hline
 $T_c$, & $\Delta _{L}$, & $\Delta _{S}$, & $\Delta _{L} + \Delta_S,$ &  $\varepsilon_0$, & $\frac{\varepsilon_0}{k_BT_c}$ \\
K & meV & meV  &  meV &  meV  &  \\
\hline
 50 & $11.7 \pm 1.2$ & -- & -- &  $11.0 \pm 1.7$ & 2.6 \\
\hline
 49 & $11.3 \pm 1.1$ & $2.2 \pm 0.3$ & $\approx 13.5$ & $14.8 \pm 2.2$ & 3.5 \\
\hline
 45 & $10.5 \pm 1.1$ & $2.8 \pm 0.3$ & $\approx 13.3$ &  $13.4 \pm 2.0$ & 3.5 \\
\hline
 37 & $9.2 \pm 0.9$ & -- & -- &  $10.1 \pm 1.5$ & 3.2 \\
\hline
 26 & $5.9 \pm 0.6$ & $1.5 \pm 0.2$ & $\approx 7.4$ &  $7.4 \pm 1.1$ & 3.3 \\
 \hline
\end{tabular}
\label{table}
\end{table}

\acknowledgments
We thank V.M. Pudalov, and H. Kontani for fruitful discussions. The work was supported by RFBR grant 17-02-00805-a. KSA acknowledges RSF grant 16-42-01100. The research has been partly done using the research equipment of the Shared facility Center at LPI.

\end{document}